\documentstyle[12pt]{article}
\begin{document}
\centerline{\bf{Second virial coefficient for a d-dimensional
Lennard-Jones (2n-n) system}}
\vskip .3in

\centerline{M.L. Glasser}

\centerline{Department of Physics and Clarkson Center for Statistical Physics}

\centerline{Clarkson University}

\centerline{Potsdam, NY 13699-5820 (USA)}

\vskip 1in
\begin{quote}

This note examines the  second virial
coefficient for an imperfect gas subject to a 2n-n interparticle
potential in any dimension $d$ between $0$ and $n$.
A compact analytic expression is presented for this quantity
which shows that, apart from a numerical factor, its temperature
dependence is a universal function parameterized by $d/n$.
\vskip 1in

\noindent
PACS:05.70.Ce, 02.30.+g

\vskip .1in
\noindent
Keywords: Lennard-Jones potential, Virial coefficient.

\end{quote}
\vfill\eject

The virial expansion has played an important role in studying imperfect
gases and has recently been invoked for the examination of systems,
such as alkanes absorbed at interfaces [1], in two dimensions. It is
therefore of some interest to look at the dimensional dependence of
the coefficients. In this note an analytic expression is presented
for the second virial coefficient in the case of the Lennard-Jones
2n-n potential
$$V(r)=4\epsilon[(\frac{\sigma}{r})^{2n}-(\frac{\sigma}{r})^n] \eqno(1)$$
for spatial dimension $0<d<n$, where it is defined.

The second virial coefficient is given by
$$B_d(T)=\frac{1}{2}\int d^dr[1-exp(-V(r)/kT)].\eqno(2)$$
Since $V(r)$ is spherically symmentric, the angular integrals in
hyperspherical coordinates can be carried out to give
$$B_d(T)=\frac{\pi^{d/2}}{\Gamma(d/2)}\int_0^{\infty}r^{d-1}[1-
e^{-V(r)/kT}]dr.\eqno(3)$$
After scaling out $\sigma$, integration by parts, and an elementary
change of integration variable, (3) reduces to
$$B_d(T)=C_d[2I_{1-d/6}-I_{-d/6}]\eqno(4)$$
where
$$a=4\epsilon/kT$$
$$C_d=\frac{\pi^{d/2}\sigma^d}{d\Gamma(d/2)}\eqno(5)$$
$$I_{\nu}=a\int_0^{\infty}x^{\nu}e^{-a(x^2-x)}dx.$$
The latter integral does not appear to have been tabulated, but is
easly evaluated as follows: Replace $exp(ax)$ by its Taylor series
and integrate term-by-term using Euler's gamma function integral to
get an infinite series summed with respect to $n$, say. Then break
this up into even ($n=2k$) and odd ($n=2k+1$) terms. In these series
replace $n!$ by $4^kk!(1/2)_k$ and $4^kk!(3/2)_k$, respectively.
The resulting series are seen to be Kummer functions[2] and we obtain
$$I_{\nu}=\frac{1}{2}a^{-(\nu-1)/2}[\Gamma(\frac{\nu+1}{2})
M(\frac{\nu+1}{2},\frac{1}{2},\frac{a}{4})+$$
$$\sqrt{a}\Gamma(\frac{\nu}{2}+1)M(\frac{\nu}{2}+1,\frac{3}{2},\frac{a}{4})
].\eqno(6)$$
Therefore, for $0<d<n$, in terms of the dimensionless temperature $\tau=4/a=
kT/\epsilon$, and $\rho=d/2n$
$$B_{d,n}(\tau)=-\frac{\sigma^d\pi^{d/2}2^{d/n}}{2n\Gamma(d/2)}
\tau^{-\rho}[\Gamma(-\rho)M(-\rho,\frac{1}{2},\tau^{-1})+$$
$$2\tau^{-1/2}\Gamma(\frac{1}{2}-\rho)M(\frac{1}{2}-\rho,
\frac{3}{2},\tau^{-1})]=\eqno(7)$$
$$\frac{\sigma^d\pi^{d/2}2^{2\rho}}{n\Gamma(d/2)}\tau^{-\rho}e^{1/\tau}[
\Gamma(-\rho)M(\rho+\frac{1}{2},\frac{1}{2},-1/\tau)+$$
$$\frac{2}{\sqrt{\tau}}\Gamma(\frac{1}{2}-\rho)M(\rho+1,\frac{3}{2},-1/\tau)]=$$
$$\frac{\sigma^d\pi^{(d-1)/2}2^{\rho}}{n\Gamma(d/2)}\Gamma(-\rho)\Gamma(
\frac{1}{2}-\rho)D_{2\rho}(-\sqrt{2/\tau}).$$

In deriving (7) the identities
$$M(1-\frac{d}{2n},\frac{1}{2},z)-2zM(1-\frac{d}{2n},\frac{3}{2},z)=
M(-\frac{d}{2n},\frac{1}{2},z)$$
$$(1-\frac{d}{n})M(\frac{3}{2}-\frac{d}{2n},\frac{3}{2},z)-M(\frac{1}{2}-
\frac{d}{2n},\frac{1}{2},z)=-\frac{d}{n}M(\frac{1}{2}-\frac{d}{2n},
\frac{3}{2},z)\eqno(8)$$
and Kummer's first identity 
have been used. The last expression in (7) is particularly attractive, but
as indicated below, the parabolic cylinder function can be problematic and
the expressions using Kummer functions are probably better for numerical
work. From Eq.(7) we have the interesting fact that, apart from a constant
prefactor, the scaled temperature dependence of the second virial
coefficient is a universal function of $d/n$.
Thus, the  Boyle temperature
( at which the second virial coefficient vanishes),  is given by the
zeros of the parabolic cylinder function. Listed below are the
Boyle temperatures for the L-J 12-6
potential in various dimensions:

$$\begin{array}{cccc}
d&\tau_B&d&\tau_B\\
0.5&0.45812&3.5&5.35369\\
1&0.72701&4&9.01642\\
1.5&1.07421&4.5&17.1531\\
2&1.56031&5&41.0597\\
2.5&2.28098&5.5&173.8854\\
3&3.41793&&

\end{array}$$
The Boyle temperature is therefore also universal with respect to $d/n$,
so this table lists the Boyle temperatures for $d/n=1/12,\dots 11/12$.

Finally, we present the Joule-Thompson coefficient $J_{d,n}(\tau)=\tau
\partial B_{d,n}(\tau)/\partial\tau-B_{d,n}(\tau)$.
$$J_{d,n}(\tau)=\frac{\sigma^d\pi^{d/2}2^{d/2n}}{6n\Gamma(d/2)}\tau^{-d/2n}\{
3\Gamma(-\frac{d}{2n})[(1+\frac{d}{2n})M(-\frac{d}{2n},\frac{1}{2},\tau^{-1})-$$
$$\frac{2}{\tau}\frac{d}{2n}M(1-\frac{d}{2n},\frac{3}{2},\tau^{-1})]+$$
$$2\tau^{-1/2}\Gamma(\frac{1}{2}-\frac{d}{2n})[3(\frac{3}{2}+
\frac{d}{2n})M(\frac{1}{2}-\frac{d}{2n}),\frac{3}{2},\tau^{-1})
+\frac{2}{\tau}(\frac{1}{2}-\frac{d}{2n})M(\frac{3}{2}-\frac{d}{2n},\frac{5}{2},
\tau^{-1})]\}.\eqno(12)$$

A similar calculation for the case $d=3$ was carried out a number of years
ago by Garrett [4] who expressed his result in terms of a sum of parabolic
cylinder functions $D_{\nu}(-z)$. This function appears to be an orphan
among special functions, in that there is very little information
about it in standard references such as [3]. What there is concerns
the function $D_{\nu}(z)$. In spite of the similarity in notation,
the two are independent solutions of the parabolic cylinder equation and
their relationship is not a simple one. For example, for real z,

$$D_{1/2}(z)=\frac{1}{\sqrt{\pi}}(z/2)^{3/2}[K_{\frac{1}{4}}(z^2/4)+
K_{\frac{3}{4}}(z^2/4)]\eqno(13)$$
while
$$D_{1/2}(-z)=\frac{\sqrt{\pi}}{4}z^{3/2}[I_{\frac{3}{4}}(z^2/4)+
I_{-\frac{3}{4}}(z^2/4)-I_{\frac{1}{4}}(z^2/4)-I_{-\frac{1}{4}}(z^2/4)].
\eqno(14)$$
On the other hand, the Kummer function used in (7)
(also denoted
$M(a,b,c)=\;_1F_1(a;b;c)$) is standard in mathematical packages and does not
suffer from this problem. For $d=3$ (7) is consistent with Garrett's
formula. Recently, Vargas et al.[5] have treated $B_{3,6}$ by an improvement of
Garrett's approach and have expressed the virial coefficient in terms
of modified Bessel functions of fractional order. 
By making use of (13) and (14), (7) can be transformed into  their expression.
Finally, the method used here for the $2n-n$
potential: series expansion, decomposition of the sum into relatively prime
summands and the use of the Gauss multiplication formula, can be
applied to the general $m-n$ potential and leads to a closed form
expression as a sum of $lcm(m,n)/n$ hypergeometric functions. Thus,
for the $12-5$ potential considered in [1], the second virial
coefficient was expressed as a sum of 12 Kummer functions, which
proved to be of little practical value.

\vfill\eject
\centerline{\bf{References}}
\vskip .1in

\noindent
[1]. B.A. Pethica, M.L. Glasser and J. Mingins, J. Colloid and Interface
Sci. {\bf{81}}, 41 (1981).

\vskip .1in
\noindent
[2]. A. Lou and B.A. Pethica, Langmuir {\bf{13}}, 4933 (1997).

\vskip .1in
\noindent
[3]. {\bf{Handbook of Mathematical Functions}},[Eds. M. Abramowitz and I.
Stegun, NBS Applied Math. Ser. 55, Washington (1964)].

\vskip .1in
\noindent
[4]. A.J.M. Garrett, J.Phys. A:Math. Gen.,{\bf{13}}, 379 (1980).

\vskip .1in
\noindent
[5]. P. Vargas, E. Mu\~nos and L. Rodriguez, Physica A{\bf{290}}, 92 (2001).

\end{document}